\documentclass{jfm}
\usepackage{epsf}

\usepackage{graphicx}
\usepackage{natbib}

\ifCUPmtlplainloaded \else
  \checkfont{eurm10}
  \iffontfound
    \IfFileExists{upmath.sty}
      {\typeout{^^JFound AMS Euler Roman fonts on the system,
                   using the 'upmath' package.^^J}%
       \usepackage{upmath}}
      {\typeout{^^JFound AMS Euler Roman fonts on the system, but you
                   dont seem to have the}%
       \typeout{'upmath' package installed. JFM.cls can take advantage
                 of these fonts,^^Jif you use 'upmath' package.^^J}%
      }
  \else
  \fi
\fi

\ifCUPmtlplainloaded \else
  \checkfont{msam10}
  \iffontfound
    \IfFileExists{amssymb.sty}
      {\typeout{^^JFound AMS Symbol fonts on the system, using the
                'amssymb' package.^^J}%
       \usepackage{amssymb}%
       \let\le=\leqslant  
         
      }{}
  \fi
\fi

\ifCUPmtlplainloaded \else
  \IfFileExists{amsbsy.sty}
    {\typeout{^^JFound the 'amsbsy' package on the system, using it.^^J}%
     \usepackage{amsbsy}}
    {\providecommand\boldsymbol[1]{\mbox{\boldmath $##1$}}}
\fi

\newcommand{\ba}{\mbox{\boldmath$a$}}
\newcommand{\be}{\mbox{\boldmath$e$}}
\newcommand{\bn}{\mbox{\boldmath$n$}}
\newcommand{\br}{\mbox{\boldmath$r$}}
\newcommand{\bu}{\mbox{\boldmath$u$}}
\newcommand{\tildeba}{\mbox{\boldmath$\tilde a$}}
\newcommand{\tildebu}{\mbox{\boldmath$\tilde u$}}

\providecommand\bnabla{\boldsymbol{\nabla}}
\providecommand\bcdot{\boldsymbol{\cdot}}

\newsavebox{\astrutbox}
\sbox{\astrutbox}{\rule[-5pt]{0pt}{20pt}}

\title[Wave attractors and tidal disturbances]
{Wave attractors and the asymptotic dissipation rate of tidal disturbances}

\author[G. I. Ogilvie]
{G\ls O\ls R\ls D\ls O\ls N\ns I.\ns O\ls G\ls I\ls L\ls V\ls I\ls E}

\affiliation{$^1$Department of Applied Mathematics and Theoretical Physics,
University of Cambridge, Centre for Mathematical Sciences, Wilberforce Road,
Cambridge CB3 0WA, UK\\[\affilskip]
$^2$Institute of Astronomy, University of Cambridge, Madingley Road,
Cambridge CB3 0HA, UK}

\pubyear{}
\volume{}
\pagerange{}
\date{10th March 2005 and in revised form 13th June 2005}
\begin{document}

\maketitle

\begin{abstract}
  Linear waves in bounded inviscid fluids do not generally form normal
  modes with regular eigenfunctions.  Examples are provided by
  inertial waves in a rotating fluid contained in a spherical annulus,
  and internal gravity waves in a stratified fluid contained in a tank
  with a non-rectangular cross-section.  For wave frequencies in the
  ranges of interest, the inviscid linearized equations are spatially
  hyperbolic and their characteristic rays are typically focused on to
  wave attractors.  When these systems experience periodic forcing,
  for example of tidal origin, the response of the fluid can become
  localized in the neighbourhood of a wave attractor.  In this paper I
  define a prototypical problem of this form and construct
  analytically the long-term response to a periodic body force in the
  asymptotic limit of small viscosity.  The vorticity of the fluid is
  localized in a detached shear layer close to the wave attractor in
  such a way that the total rate of dissipation of energy is
  asymptotically independent of the viscosity.  I further demonstrate
  that the same asymptotic dissipation rate is obtained if a
  non-viscous damping force is substituted for the Navier--Stokes
  viscosity.  I discuss the application of these results to the
  problem of tidal forcing in giant planets and stars, where the
  excitation and dissipation of inertial waves may make a dominant, or
  at least important, contribution to the orbital and spin evolution.
\end{abstract}

\section{Introduction}
\label{s:introduction}

Geophysics and astrophysics give rise to problems involving waves in
axisymmetric rotating fluid bodies.  The fluid outer core of the Earth
supports inertial waves, for which the Coriolis effect provides the
restoring force.  Such waves have been tentatively identified in
gravimetric data and related to theoretical work as well as laboratory
experiments (e.g. Aldridge \& Lumb 1987).  Helioseismology uses the
observed frequencies of solar oscillations to deduce the internal
structure and rotation of the Sun, while asteroseismology applies
related techniques to distant stars (e.g. Christensen-Dalsgaard 2002).
Waves in astrophysical accretion discs have been studied because of
their important role in the interaction of a planet or other satellite
with the disc in which it forms (e.g. Goldreich \& Tremaine 1980), and
also in an attempt to explain observed quasi-periodic oscillations
from accreting white dwarfs, neutron stars and black holes (e.g. Kato
2001).  Some of these problems involve free modes of oscillation that
may grow through mechanisms of overstability or may be excited by
turbulent noise.  Tidal interactions between orbiting and spinning
bodies, however, involve an almost strictly periodic forcing of waves.

After the linearized wave equations governing the dynamics of the
fluid are reduced by Fourier analysis in time and azimuth, there
results a two-dimensional problem for the spatial structure of the
wave in the meridional plane (see Appendix~\ref{a:character}).  In the
absence of dissipative effects such as viscosity, this problem may
have either elliptic or hyperbolic character depending on the
frequency of the wave.  A mixed-type problem may also occur if the
character of the equations changes from one part of the body to
another.  Typically, hyperbolic character occurs when the wave
frequency (Doppler-shifted, in the case of a non-axisymmetric wave,
into the frame rotating with the local fluid) is small, in the range
of inertial and gravity waves rather than acoustic waves.  When a
hyperbolic system of equations is posed in a finite region with
physical boundary conditions on a closed surface, rather than the
mathematically canonical Cauchy boundary conditions on an open
surface, the problem can be regarded as ill-posed and generally does
not possess a regular solution.  This property presents a serious
difficulty when one seeks the response of a body to low-frequency
tidal forcing.\footnote{In accretion discs, where the systematic shear
causes a strong Doppler shift of the frequency of non-axisymmetric
waves, the equations typically have hyperbolic character in the
vicinity of the corotation resonance.}

It is possible to sidestep this issue in certain problems of special
symmetry where a separation of variables can be applied.  For example,
in a spherically symmetric non-rotating star the linearized equations
can be projected on to spherical harmonics and the problem reduced to
a system of ordinary differential equations in the radial direction.
Most treatments of tides in stars and giant planets have used this
approach (e.g. Zahn 1970), or have included the Coriolis force within
the so-called `traditional approximation' (in which only the radial
component of the angular velocity is considered), which also permits a
separation of variables (e.g. Ioannou \& Lindzen 1993; Savonije \&
Witte 2002).  The simplest problems involving inertial waves,
featuring an incompressible fluid in a full spherical, spheroidal or
cylindrical container, can also be solved by separation of variables
and appear to possess complete sets of normal modes with regular
eigenfunctions, even in the absence of viscosity (e.g. Greenspan 1968;
Zhang, Liao \& Earnshaw 2004).

Unfortunately the approximation of a non-rotating star is rarely valid
in problems of tidal forcing.  The tidal frequencies are linear
combinations of the orbital and spin frequencies with small integer
coefficients, apart from small corrections due to any precessional
effects.  In most problems of interest, the frequencies of the most
important tidal components are smaller in magnitude than twice the
spin frequency of the body, and the Coriolis force cannot be neglected
(Ogilvie \& Lin 2004).  The traditional approximation is valid only in
stably stratified regions where the tidal frequency is much less than
the buoyancy frequency, and does not apply in convective regions of
stars or giant planets where the buoyancy frequency is essentially
zero.  In such regions the wave equations are hyperbolic in character,
describing pure inertial waves, and cannot be solved by separation of
variables.  Numerical solutions of the two-dimensional problem were
obtained by Savonije, Papaloizou \& Alberts (1995) and Savonije \&
Papaloizou (1997) in the case of an early-type star with a small
convective core and, more recently, by Ogilvie \& Lin (2004) in the
case of a giant planet with an extended convective region.

Numerical studies have revealed the intricate structure of inertial
waves in an incompressible rotating fluid contained in a spherical
annulus (Hollerbach \& Kerswell 1995; Rieutord \& Valdettaro 1997).
In order to find normal modes it is necessary to include a viscosity
so that the problem becomes of elliptic character and is
mathematically regularized.  As the viscosity tends to zero the
eigenfunctions become increasing localized in the neighbourhood of
singular linear structures known as wave attractors, as described in
greater detail by Rieutord, Georgeot \& Valdettaro (2001) and
Rieutord, Valdettaro \& Georgeot (2002).  A wave attractor can be
understood as a limit cycle towards which the characteristic rays of
the inviscid wave equation are focused as they reflect repeatedly from
the boundaries of the container.  Unlike the case of a full sphere, it
appears that inertial waves in a spherical annulus do not possess
regular eigenfunctions in the absence of viscosity, apart from the
exceptional `r modes' or `toroidal modes' which involve no radial
motion.  The presence of an inner core introduces a complexity into
the reflection patterns of the characteristic rays and causes them to
be focused on to wave attractors.  The existence and importance of
these closed ray circuits in spherical shells have been known for some
time (Stern 1963; Bretherton 1964) and many features of this problem
were understood before the advent of high-resolution numerical
calculations (Stewartson \& Rickard 1969; Stewartson 1972).

Wave attractors have also been studied for internal gravity waves.  In
a notable experiment, Maas et al. (1997) applied a periodic forcing to
a narrow tank containing a stratified salt solution.  For wave
frequencies smaller in magnitude than the uniform buoyancy frequency
of the fluid, the characteristic rays of the inviscid wave equation
are straight lines with a definite angle of inclination depending on
the frequency.  In an upright rectangular tank the rays would
propagate around the container ergodically or, for special frequencies
related to the rational numbers, would close after a finite number of
reflections.  By making one of the side-walls sloping, Maas et
al. (1997) introduced a wave attractor, in this case an inclined
rectangle, into the problem (Maas \& Lam 1995).  Their experiment
indicates that the forced response is localized in the neighbourhood
of the attractor, and this concentration ultimately leads to secondary
effects such as mixing.  Inertial waves in a rotating tank of similar
shape have also been studied by Maas (2001) and Manders \& Maas
(2003).

There is a close analogy between these problems.  The relation between
the tank with the sloping side-wall and the upright rectangular tank
is similar to the relation between the spherical annulus and the full
sphere.  In each case rays in the geometrically simpler container
(rectangular tank or full sphere) propagate ergodically or, for
special frequencies, form periodic orbits.  The inviscid wave equation
admits regular eigenfunctions for a countable set of frequencies,
which may or may not be those for which the rays form periodic orbits.
In the geometrically more complicated, but more generic, containers
the rays are focused on to one or more wave attractor for almost all
wave frequencies, and the inviscid wave equation does not admit
regular eigenfunctions.

In problems of tidal forcing it is of great interest to know how the
total dissipation rate varies with the forcing frequency, as this
determines the rate of secular evolution of tidally interacting
systems.  When the viscosity is small, systems that possess regular
inviscid normal modes can be expected to exhibit a strong resonant
amplification of the dissipation rate in the vicinity of the
eigenfrequencies, but very small dissipation elsewhere.  Although the
eigenfrequencies may be everywhere dense in some interval, the smooth
forcing will have a significant overlap only with a few of the
lowest-order modes.  In contrast, wave attractors have a structural
stability and exist in intervals of frequency that depend on the
geometry rather than the viscosity.  Systems possessing wave
attractors can be expected to exhibit a richer response with
significant dissipation occurring over extended ranges of frequency.

The purpose of this paper is to analyse the linear response of such
systems to periodic forcing in the low-viscosity limit relevant to the
geophysical and astrophysical applications.  I work with simplified
model problems but also explain how they are related to the original
systems.  The main result is a validation of the conjecture of Ogilvie
\& Lin (2004) that, in problems where a wave attractor occurs, the
total dissipation rate tends to a non-zero value that is independent
of both the magnitude and the form of the small-scale damping process
of the waves.  This outcome differs significantly from problems that
possess regular inviscid normal modes and is of primary importance for
tidally interacting systems.  The analysis presented here provides a
method of calculating the asymptotic dissipation rate and also
describes the spatial form of the forced disturbance.

The remainder of this paper is organized as follows.  In
\S\ref{s:maas} I briefly describe the problem of forced internal
gravity waves as a definite example of the system under consideration.
I define a prototypical problem in \S\ref{s:prototypical} and
construct the asymptotic solution in the limit of small viscosity.  In
\S\ref{s:non-viscous} I adapt the analysis to a related problem in
which the damping mechanism is not of a viscous nature.  I present the
results of direct numerical calculations in \S\ref{s:numerical} and
compare these with the asymptotic theory.  A concluding discussion is
found in \S\ref{s:discussion}.

\section{Forced internal gravity waves}
\label{s:maas}

In this section I outline the preliminary analysis for forced internal
gravity waves in a narrow tank.  This is essentially identical to the
problem studied by Maas et al. (1997) except that a vortical body
force is assumed rather than a parametric forcing.

Consider a fluid initially at rest in a gravitational field
$-g\,\be_z$ and with a uniform temperature gradient $\beta\,\be_z$,
where $(x,y,z)$ are Cartesian coordinates.  In the Boussinesq
approximation the linearized equations for the velocity $\bu$ and the
temperature perturbation $\theta$ are (Chandrasekhar 1961)
\begin{equation}
  \frac{\partial\bu}{\partial t}=-\bnabla\varpi+g\alpha\theta\,\be_z+
  \nu\nabla^2\bu+\ba,
\end{equation}
\begin{equation}
  \frac{\partial\theta}{\partial t}=-\beta u_z+\kappa\nabla^2\theta,
\end{equation}
\begin{equation}
  \bnabla\bcdot\bu=0,
\end{equation}
where $\varpi$ is the pressure perturbation divided by the reference
density, $\alpha$ is the coefficient of expansion, $\nu$ is the
kinematic viscosity and $\kappa$ is the thermal diffusivity.  The
fluid is excited by an external body force $\ba$ per unit mass.

Consider a two-dimensional problem, approximately representative of
the situation in a narrow tank, in which $u_y=a_y=0$ and all
quantities are independent of $y$.  The velocity is described by a
streamfunction $\psi$ such that
\begin{equation}
  u_x=-\frac{\partial\psi}{\partial z},\qquad
  u_z=\frac{\partial\psi}{\partial x}.
\end{equation}
I focus on the limit of large Prandtl number in which $\kappa$ can be
neglected (except in thermal boundary layers, which are of negligible
thickness compared to the viscous boundary layers).  In reality, the
Prandtl number of water exceeds $10$ for temperatures below about
$10^\circ{\rm C}$, so this limit may be of some relevance for
laboratory experiments or for terrestrial tides confined in oceanic
basins.

When considering the long-term response to a periodic force with
(real) angular frequency $\omega$, all perturbation quantities may be
assumed to have the form
\begin{equation}
  \bu={\rm Re}\left[\tildebu(x,z)\,{\rm e}^{-{\rm i}\omega t}\right],
\end{equation}
etc.  Eliminating $\tilde\varpi$ and $\tilde\theta$, one obtains the
linearized vorticity equation in the form
\begin{equation}
  {\rm i}\omega\nabla^2\tilde\psi=\frac{{\rm i}N^2}{\omega}
  \frac{\partial^2\tilde\psi}{\partial x^2}-\nu\nabla^4\tilde\psi+f,
\label{eq_maas}
\end{equation}
where $N^2=g\alpha\beta$ is the square of the buoyancy frequency and
\begin{equation}
  f=\frac{\partial\tilde a_x}{\partial z}-
  \frac{\partial\tilde a_z}{\partial x}
\end{equation}
is the $y$-component (and only non-vanishing component) of the
vorticity forcing\footnote{It may be noted that tidal forcing always
derives from a potential and therefore has no curl.  Nevertheless, the
dynamical tide in a body of non-uniform density, or one with a free
surface, does experience a vortical effective forcing as explained in
Appendix~\ref{a:vortical}.} $\bnabla\times\tildeba$.

With rigid walls the boundary conditions are
\begin{equation}
  \tilde\psi=0,\qquad
  \bn\bcdot\bnabla\tilde\psi=0,
\end{equation}
where $\bn$ is the outward normal vector on the boundary.  The
time-averaged dissipation rate (per unit length in the $y$-direction)
is
\begin{equation}
  D=\frac{1}{2}\int\!\!\int\nu|\nabla^2\tilde\psi|^2\,{\rm d}x\,{\rm d}z,
\label{d1}
\end{equation}
where the integral extends over the area of the container in the
$xz$-plane.  Using equation (\ref{eq_maas}) and the boundary
conditions one can relate the dissipation to an overlap integral
between the vorticity forcing and the streamfunction in the form
\begin{equation}
  D=\frac{1}{2}\,{\rm Re}\int\!\!\int\tilde\psi^*f\,{\rm d}x\,{\rm d}z.
\label{d2}
\end{equation}
 
The spectrum of internal gravity waves lies in the interval
$-N<\omega<N$.  In this range of frequencies the inviscid version of
equation (\ref{eq_maas}), with $\nu=0$, involves a hyperbolic operator
\begin{equation}
  \omega^2\nabla^2-N^2\frac{\partial^2}{\partial x^2}=
  \omega^2\frac{\partial^2}{\partial z^2}-
  (N^2-\omega^2)\frac{\partial^2}{\partial x^2}.
\end{equation}
The characteristics of the inviscid equation are straight lines with
slopes $\pm\omega(N^2-\omega^2)^{-1/2}$ that depend on the wave
frequency.  The occurrence of wave attractors in containers of various
shapes has been illustrated by Maas et al. (1997).

\section{A prototypical problem and its asymptotic solution}
\label{s:prototypical}

\subsection{Definition of the problem}

In this section I slightly abstract the above problem and define a
prototypical model problem that is very closely related but marginally
simpler to solve.  Consider the equation
\begin{equation}
  {\rm i}\frac{\partial^2\psi}{\partial x\partial y}+\epsilon^3\nabla^4\psi=f,
\label{eq_del4}
\end{equation}
for an unknown function $\psi(x,y)$ in a domain ${\cal D}\subset{\bf
R}^2$, with boundary conditions
\begin{equation}
  \psi=0,\qquad
  \bn\bcdot\bnabla\psi=0\qquad
  \hbox{on $\partial{\cal D}$}.
\label{bc}
\end{equation}
Here $\epsilon\ll1$ is a small dimensionless parameter, $f(x,y)$ is a
given complex-valued function and $\bn$ is the outward normal vector
on $\partial{\cal D}$.

As suggested by \S\ref{s:maas}, one can think of $\psi$ as the
streamfunction of a two-dimensional flow of an incompressible fluid
subject to no-slip boundary conditions.  The viscosity of the fluid is
proportional to $\epsilon^3$ and the curl of the body force is
proportional to $f$.  The limit $\epsilon\ll1$ corresponds to the
Reynolds number being large.  The problem is linearized and a harmonic
time-dependence of all quantities has been assumed.  The fluid has a
restoring force such as buoyancy or the Coriolis force.  In fact the
inviscid part of the problem of forced internal gravity waves
considered above can be rendered exactly in this form by a linear
transformation of the coordinates, depending on the wave frequency,
that maps the sloping characteristics on to horizontal and vertical
lines.  In the process the viscous $\nabla^4$ operator is slightly
modified, although this has essentially no practical consequences.

Consider the inviscid version of equation (\ref{eq_del4}), obtained by
setting $\epsilon=0$ and abandoning the no-slip boundary condition
$\bn\bcdot\bnabla\psi=0$.  The inviscid equation is hyperbolic and its
characteristics are horizontal and vertical lines.  In general, the
inviscid problem is mathematically ill-posed and does not possess a
regular solution.  Instead, one must consider the limit $\epsilon\to0$
of the viscous problem.

The time-averaged energy dissipation rate may be defined as
\begin{equation}
  D=\frac{1}{2}\int_{\cal D}\epsilon^3|\nabla^2\psi|^2\,{\rm d}A.
\end{equation}
Using equation (\ref{eq_del4}) and the boundary conditions (\ref{bc}) one
can relate the dissipation to an overlap integral between the
vorticity forcing and the streamfunction in the form
\begin{equation}
  D=\frac{1}{2}{\rm Re}\int_{\cal D}\psi^*f\,{\rm d}A.
\end{equation}
The problem at hand is to determine how $D$ depends on $\epsilon$ in
the limit $\epsilon\to0$ (and also on $f$ and ${\cal D}$ if these are
varied).

\subsection{Ray circuits and the wave attractor}
\label{s:circuits}

A \emph{ray segment} consists of a horizontal or vertical line segment
connecting two points of $\partial{\cal D}$ (\emph{vertices}) and
lying wholly in ${\cal D}$.  A \emph{ray circuit} consists of a
connected set of consecutive ray segments and represents the
propagation of a wave characteristic around the container.  Typically
a ray circuit will be of infinite length.  Exceptionally, a circuit
that closes on itself after a finite number of reflections is a
\emph{periodic orbit}.

\begin{figure}
  \centerline{\epsfysize6cm\epsfbox{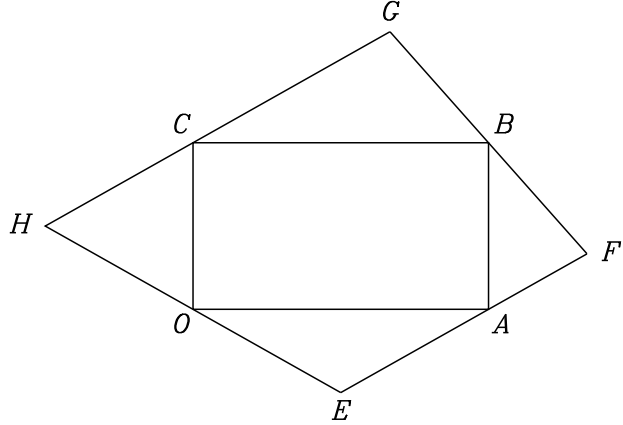}}
  \caption{Example of a quadrilateral domain containing a unique wave
     attractor.}
\label{f:quadrilateral}
\end{figure}

In principle, the domain ${\cal D}$ may contain any number of periodic
orbits.  I consider the case in which ${\cal D}$ contains exactly one
periodic orbit and the orbit is simple (i.e. a rectangle).  Without
loss of generality one may place the origin of coordinates at the
bottom left-hand corner of the rectangle so that the corners $OABC$
are at $(0,0)$, $(X,0)$, $(X,Y)$ and $(0,Y)$ with $X,Y>0$.  I label
the segments 1, 2, 3,~4 in the order $OA$, $AB$, $BC$, $CO$.  Let the
slopes of the boundary at the four corners be $s_A=t_1$,
$s_B=-t_2^{-1}$, $s_C=t_3$ and $s_O=-t_4^{-1}$.  Then $t_j>0$ is the
tangent of the angle between the `incident' ray (in the positive sense
$OABC$) and the boundary at the end of segment~$j$.

For rays close to the periodic orbit, the boundary may be represented
locally by linear approximations
\begin{eqnarray}
  y&\approx&t_1(x-X)\phantom{-t_1^{-1}x+Y}\qquad\hbox{near
  $A$},\nonumber\\
  y&\approx&-t_2^{-1}(x-X)+Y\phantom{t_2x}\qquad\hbox{near
  $B$},\nonumber\\
  y&\approx&t_3x+Y\phantom{-t_3^{-1}(x-X)}\qquad\hbox{near
  $C$},\nonumber\\
  y&\approx&-t_4^{-1}x\phantom{t_4(x-X)+Y}\qquad\hbox{near $O$}.
\end{eqnarray}
The local approximate mapping of vertices defined by the ray
propagation (in the positive sense $OABC$) is then
\begin{eqnarray}
  (-t_4y,y)&\mapsto&(t_1^{-1}y+X,y)\nonumber\\
  &\mapsto&(t_1^{-1}y+X,-t_1^{-1}t_2^{-1}y+Y)\nonumber\\
  &\mapsto&(-t_1^{-1}t_2^{-1}t_3^{-1}y,-t_1^{-1}t_2^{-1}y+Y)\nonumber\\
  &\mapsto&(-t_1^{-1}t_2^{-1}t_3^{-1}y,t_1^{-1}t_2^{-1}t_3^{-1}t_4^{-1}y).
\end{eqnarray}
On each loop, therefore,
\begin{equation}
  (-t_4y,y)\mapsto\alpha^{-1}(-t_4y,y),\qquad
  \alpha=t_1t_2t_3t_4.
\end{equation}
It may be expected that $\alpha\ne1$ in a typical asymmetric
container.  The periodic orbit is then attracting, and focuses rays
propagating either in a positive sense (if $\alpha>1$) or in a
negative sense (if $0<\alpha<1$).  I assume without loss of generality
that $\alpha>1$ (the case $0<\alpha<1$ can be obtained by a reflection
of the problem).  The focusing direction is then $OABC$ and $\alpha$
is the focusing power of the attractor.

A particularly useful illustrative example is provided by a
quadrilateral container (figure~\ref{f:quadrilateral}).  The
quadrilateral is defined uniquely by specifying the rectangle $OABC$
and the four tangents $t_j$.  It can be shown that the wave attractor
is unique.  All ray circuits other than the attractor itself converge
towards the attractor at either end.  I assume that this property
holds in the domain ${\cal D}$.

The attractor divides the boundary into two zones, one consisting of
$OEA$ together with $BGC$ and the other consisting of $AFB$ together
with $CHO$.  Ray circuits other than the attractor itself have all
their vertices in one zone or the other.  Therefore the attractor has
two `sides' to it, and any ray circuit converges towards the attractor
from the same side at either end.

For rays near the attractor one can measure the distance from the
attractor using the $y$-coordinates of the appropriate vertices (the
ones near $O$).  The entire bundle of rays associated with one side of
the attractor (e.g. the `positive' side $y>0$) can be labelled using a
variable $\theta^+\in[0,1)$ defined as follows.  One starts a ray at a
vertex very close to $O$ and with $y$-coordinate
$y=\alpha^{-n-\theta^+}$ for some arbitrarily large positive integer
$n\gg1$.  The ray is followed in the negative sense as it spirals away
from the attractor.  Eventually it turns around and converges towards
the attractor again, having vertices very close to $O$ with
$y$-coordinates $y\sim\alpha^{-n-\Theta^+(\theta^+)}$, $n\gg1$.  This
mapping between one end of a ray circuit and the other defines a
function $\Theta^+(\theta^+)$, with $\Theta^+\in[0,1)$.

The variable $\theta^+$ is periodic in nature ($\theta^+=1$ is
identified with $\theta^+=0$) since the ray bundle wraps around
itself.  The function $\Theta^+(\theta^+)$ can be considered
topologically as a map of the circle on to itself.  Owing to the
reversibility of the ray circuit it has the self-inverting property
$\Theta^+(\Theta^+(\theta^+))=\theta^+$, which will be used later.

For the negative side $y<0$ of the attractor there is a similar
variable $\theta^-$ and a mapping $\Theta^-(\theta^-)$.  In problems
with multiple or non-simple attractors a more complicated mapping of
ray bundles is likely to exist.

\subsection{Construction of an asymptotic solution}

In the limit $\epsilon\to0$ the solution divides into three regions:
(i) an inner region localized near the wave attractor and consisting
of a detached shear layer of width $O(\epsilon)$; (ii) an outer region
consisting of most of the rest of the container; (iii) standard
viscous boundary layers of width $O(\epsilon^{3/2})$ close to the
walls.  There are also boundary layers in the corner regions close to
the vertices of the attractor.

I first simplify the problem by assuming that a particular solution
for the inviscid problem can be found, i.e. a function $\hat\psi(x,y)$
satisfying the equation
\begin{equation}
  {\rm i}\frac{\partial^2\hat\psi}{\partial x\partial y}=f
\end{equation}
without regard to boundary conditions.  (This can be obtained by an
indefinite integration of $-{\rm i}f$ with respect to $x$ and $y$.)
The desired solution is then $\psi=\hat\psi+\tilde\psi$ where
$\tilde\psi$ satisfies the equation
\begin{equation}
  {\rm i}\frac{\partial^2\tilde\psi}{\partial x\partial y}+
  \epsilon^3\nabla^4\tilde\psi=-\epsilon^3\nabla^4\hat\psi
\label{eq_tilde}
\end{equation}
and the boundary conditions
\begin{equation}
  \tilde\psi=-\hat\psi,\qquad
  \bn\bcdot\bnabla\tilde\psi=-\bn\bcdot\bnabla\hat\psi.
\end{equation}
Note that $f$ and $\hat\psi$ are supposed to be smooth and do not have
any fine structure associated with boundary layers or shear layers.
Therefore the right-hand side of equation (\ref{eq_tilde}) is small
everywhere and will be unimportant in constructing the leading-order
asymptotic solution.  The inhomogeneity has effectively been
transferred from the differential equation to the boundary conditions.

\subsection{The outer solution}

The asymptotic outer solution is simply of the form
\begin{equation}
  \tilde\psi\sim\tilde\psi^{\rm(out)}(x,y).
\end{equation}
It satisfies the inviscid problem
\begin{equation}
  {\rm i}\frac{\partial^2\tilde\psi^{\rm(out)}}{\partial x\partial y}=0
\label{eq_outer}
\end{equation}
with the boundary condition
\begin{equation}
  \tilde\psi^{\rm(out)}=-\hat\psi.
\label{bc_outer}
\end{equation}
The second boundary condition is taken care of by the intervention of
a standard viscous boundary layer (region (iii) mentioned above).

The general solution of equation (\ref{eq_outer}) is
\begin{equation}
  \tilde\psi^{\rm(out)}=g(x)-h(y),
\label{sol_outer}
\end{equation}
where $g$ and $h$ are functions to be determined.  Consider two
consecutive vertices, $P$ and $Q$, say, of a ray circuit.  If they are
connected by a horizontal ray segment, they share the same value of
$y$ and therefore of $h(y)$.  It follows from equations
(\ref{bc_outer}) and (\ref{sol_outer}) that
$g(x_Q)-g(x_P)=\hat\psi_P-\hat\psi_Q$.  Similarly, for vertices
connected by a vertical ray segment,
$h(y_Q)-h(y_P)=\hat\psi_Q-\hat\psi_P$.  Since $\hat\psi$ is known, a
knowledge of $g$ (or $h$) at any vertex would be sufficient to
propagate the solution along the ray circuit.  However the given
boundary conditions do not provide such information.

Consider the ray circuit as described in \S\ref{s:circuits}.  At each
end of the circuit the ray indefinitely repeats a loop around the
attractor $OABC$.  Around each loop the value of $g$ must change by an
amount
\begin{equation}
  \delta=\hat\psi_O-\hat\psi_A+\hat\psi_B-\hat\psi_C,
\end{equation}
and the value of $h$ also changes by $\delta$.  Since $\hat\psi$ is
smooth, one need not worry about the small variation of $\hat\psi$
close to the vertices of the attractor.  In fact
\begin{equation}
  \delta=\int\!\!\int
  \frac{\partial^2\hat\psi}{\partial x\partial y}\,{\rm d}x\,{\rm d}y=
  -{\rm i}\int f\,{\rm d}A,
\end{equation}
where the integration is over the area enclosed by the attractor.
Therefore $\delta$ is independent of the choice of particular solution
$\hat\psi$.  It is also notable that if, as in \S\ref{s:maas}, $f$ is
the perpendicular component of the curl of the external force per unit
mass, then Stokes's theorem provides a simple relation between
$\delta$ and the line integral $\oint\tildeba\bcdot{\rm d}\br$ of the
force around the attractor.

It follows that $h(y)$ exhibits an increasingly fine structure and, in
general, a logarithmic divergence as $y\to0$.  The same is true of
$g(x)$ near $x=0$ and $x=X$, and of $h(y)$ near $y=Y$.  This singular
behaviour of the outer solution is the reason that an inner region, in
which viscosity regularizes the solution, is required.  For small $y$,
one has
\begin{equation}
  h(\alpha^{-1}y)-h(y)\approx\delta.
\end{equation}
Consider the positive side $y>0$ first and define the variable $\tilde
y^+=-\ln y/\ln\alpha$, which increases by $1$ on each loop.  Then
$h(y)\sim\tilde h^+(\tilde y^+)$ as $y\searrow0$, with
\begin{equation}
  \tilde h^+(\tilde y^++1)-\tilde h^+(\tilde y^+)=\delta.
\end{equation}
The general solution of this functional equation is
\begin{equation}
  \tilde h^+(\tilde y^+)=\tilde y^+\delta+H^+(\theta^+),
\end{equation}
i.e.
\begin{equation}
  h(y)\sim\left(-\frac{\ln y}{\ln\alpha}\right)\delta+H^+
  (\theta^+)\qquad
  \hbox{as $y\searrow0$},
\label{hplus_limit}
\end{equation}
where
\begin{equation}
  \theta^+=\tilde y^+\bmod 1=
  \left(-\frac{\ln y}{\ln\alpha}\right)\bmod1
\end{equation}
is the ray-labelling variable introduced previously and $H^+$ is an
arbitrary function with period $1$.  It may be expanded in a Fourier
series
\begin{equation}
  H^+(\theta^+)=\sum_{n=-\infty}^\infty H_n^+{\rm e}^{2n\pi{\rm i}\theta^+}.
\end{equation}
For the negative side of the attractor, a similar result holds, with
\begin{equation}
  h(y)\sim\left(-\frac{\ln|y|}{\ln\alpha}\right)\delta+
  H^-(\theta^-)\qquad
  \hbox{as $y\nearrow0$}
\end{equation}
and
\begin{equation}
  H^-(\theta^-)=\sum_{n=-\infty}^\infty H_n^-{\rm e}^{2n\pi{\rm i}\theta^-}.
\end{equation}

Now consider the total change in $h(y)$ around the ray circuit, from a
vertex with $y=\alpha^{-n_1-\theta^+}$, $n_1\gg1$, to a vertex with
$y\sim\alpha^{-n_2-\Theta^+(\theta^+)}$, $n_2\gg1$.  This change can
be computed numerically by summing the values of $\hat\psi$, with
appropriate signs, at the vertices visited by the ray circuit.  The
total change can be written as
\begin{equation}
  \left[n_2+\Theta^+(\theta^+)-n_1-\theta^+\right]\delta+\Delta^+(\theta^+),
\end{equation}
where $\Delta^+$ does not depend on $n_1$ or $n_2$ because to
increment either by $1$ adds a loop around the attractor at one end
and changes $h(y)$ by $\mp\delta$ at that end.  Owing to the
reversibility of the ray circuit,
$\Delta^+(\Theta^+(\theta^+))=-\Delta^+(\theta^+)$.

Connecting the asymptotic forms, equation (\ref{hplus_limit}), of the
solution at the two ends of the circuit, one finds
\begin{equation}
  H^+(\Theta^+(\theta^+))-H^+(\theta^+)=\Delta^+(\theta^+).
\end{equation}
The general solution of this functional equation is
\begin{equation}
  H^+(\theta^+)=-\frac{1}{2}\Delta^+(\theta^+)+J^+(\theta^+),
\end{equation}
where $J^+$ is an arbitrary function satisfying the symmetry condition
$J^+(\Theta^+(\theta^+))=J^+(\theta^+)$.  An analogous equation holds
for the negative side of the attractor.  There is insufficient
information to determine the solution uniquely until the equations for
$H^+$ and $H^-$ can be linked through the dynamics in the inner
region.

\subsection{The inner solution}

The inner solution is localized near the attractor and consists of
four segments.  I begin by defining coordinates parallel ($\xi$) and
perpendicular ($\eta$) to the ray in each segment:
\begin{eqnarray}
  &&\xi_1=x,\phantom{y(X-x)(Y-y)}\!\!\!\!\!\!\!\!
  \eta_1=\epsilon^{-1}y\phantom{(-x)(x-X)(Y-y)}\!\!\!\!\!\!\!\!
  \hbox{(segment $OA$)},\nonumber\\
  &&\xi_2=y,\phantom{x(X-x)(Y-y)}\!\!\!\!\!\!\!\!
  \eta_2=\epsilon^{-1}(x-X)\phantom{(-x)y(Y-y)}\!\!\!\!\!\!\!\!
  \hbox{(segment $AB$)},\nonumber\\
  &&\xi_3=(X-x),\phantom{xy(Y-y)}\!\!\!\!\!\!\!\!
  \eta_3=\epsilon^{-1}(Y-y)\phantom{(-x)y(x-X)}\!\!\!\!\!\!\!\!
  \hbox{(segment $BC$)},\nonumber\\
  &&\xi_4=Y-y,\phantom{(xy(X-x))}\!\!\!\!\!\!\!\!
  \eta_4=\epsilon^{-1}(-x)\phantom{y(x-X)(Y-y)}\!\!\!\!\!\!\!\!
  \hbox{(segment $CO$)}.
\end{eqnarray}
In segment~$j$ the parallel coordinate $\xi_j$ runs from $0$ to
$\Xi_j$, where $\Xi_1=\Xi_3=X$ and $\Xi_2=\Xi_4=Y$.  The perpendicular
coordinate $\eta_j$ is stretched to resolve the inner region of width
$O(\epsilon)$.  In segment~$j$ the asymptotic inner solution is of the
form
\begin{equation}
  \tilde\psi\sim\tilde\psi_j^{\rm(in)}(\xi_j,\eta_j)
\end{equation}
and satisfies the equation
\begin{equation}
  {\rm i}\frac{\partial^2\tilde\psi_j^{\rm(in)}}
  {\partial\xi_j\partial\eta_j}+
  \frac{\partial^4\tilde\psi_j^{\rm(in)}}{\partial\eta_j^4}=0,
\label{eq_inner}
\end{equation}
which follows from equation (\ref{eq_tilde}) under the assumed
scalings.

Part of the difficulty in solving equation (\ref{eq_inner}) consists
of connecting the four segments.  Boundary conditions apply in the
corner regions where the segments meet and the wave is reflected.
Where segment~$j$ overlaps with segment $j+1$, the asymptotic solution
consists simply of the linear superposition of
$\tilde\psi_j^{\rm(in)}$ and $\tilde\psi_{j+1}^{\rm(in)}$, because
equation (\ref{eq_inner}) is linear and homogeneous.  In the corner
region, $\xi_j$ and $\xi_{j+1}$ are essentially constant while
$\eta_j$ and $\eta_{j+1}$ vary by $O(1)$.  Therefore the corner
solution is of the form
\begin{equation}
  \tilde\psi\sim\tilde\psi_j^{\rm(in)}(\Xi_j,\eta_j)+
  \tilde\psi_{j+1}^{\rm(in)}(0,\eta_{j+1}).
\end{equation}
The location of the boundary is given by
\begin{equation}
  \eta_{j+1}=t_j^{-1}\eta_j+O(\epsilon).
\end{equation}
From the boundary condition $\tilde\psi=-\hat\psi$ one obtains the
relation
\begin{equation}
  \tilde\psi_{j+1}^{\rm(in)}(0,t_j^{-1}\eta_j)=
  -\tilde\psi_j^{\rm(in)}(\Xi_j,\eta_j)-\hat\psi_j
\label{bc_inner}
\end{equation}
which connects the solutions in consecutive segments.  (To satisfy the
second boundary condition $\bn\bcdot\bnabla\psi=0$ a thinner boundary
layer must intervene in the corner.)  At the vertex $O$ the
corresponding relation is
\begin{equation}
  \tilde\psi_1^{\rm(in)}(0,t_4^{-1}\eta_4)=
  -\tilde\psi_4^{\rm(in)}(\Xi_4,\eta_4)-\hat\psi_4.
\end{equation}

I now define a connected inner solution by concatenating the
solutions in the various segments.  To connect the perpendicular
coordinates smoothly, let
\begin{equation}
  \eta=\cases{
  \eta_1&in segment $OA$,\cr
  t_1\eta_2&in segment $AB$,\cr
  t_1t_2\eta_3&in segment $BC$,\cr
  t_1t_2t_3\eta_4&in segment $CO$,}
\end{equation}
i.e. $\eta=f_j\eta_j$ in segment~$j$, with $f_1=1$ and
\begin{equation}
  f_j=\prod_{k=1}^{j-1}t_k,\qquad j>1.
\end{equation}
To leave equation (\ref{eq_inner}) in the same form, the parallel
coordinate $\xi_j$ must be rescaled by a factor $f_j^3$.  For a
continuous concatenation of the parallel coordinates, let
\begin{equation}
  \xi=\cases{
  \xi_1&in segment $OA$,\cr
  f_2^3\xi_2+\Xi_1&in segment $AB$,\cr
  f_3^3\xi_3+\Xi_1+f_2^3\Xi_2&in segment $BC$,\cr
  f_4^3\xi_4+\Xi_1+f_2^3\Xi_2+f_3^3\Xi_3&in segment $CO$,}
\label{xi}
\end{equation}
i.e. $\xi=\xi_1$ in segment~1 and
\begin{equation}
  \xi=f_j^3\xi_j+\sum_{k=1}^{j-1}f_k^3\Xi_k
\end{equation}
in segment $j>1$.  This coordinate runs from $0$ to $\Xi$ around the
attractor, where
\begin{equation}
  \Xi=\sum_{j=1}^4f_j^3\Xi_j=(1+t_1^3t_2^3)X+t_1^3(1+t_2^3t_3^3)Y.
\label{Xi}
\end{equation}
The solution itself, away from the corners, may then be written in the
form
\begin{equation}
  \tilde\psi\sim\tilde\psi^{\rm(in)}(\xi,\eta)
\end{equation}
with
\begin{equation}
  \tilde\psi^{\rm(in)}(\xi,\eta)=
  (-1)^{j-1}\tilde\psi_j^{\rm(in)}(\xi_j,\eta_j)+
  \sum_{k=1}^{j-1}(-1)^k\hat\psi_k
\end{equation}
in segment~$j$.  The boundary condition (\ref{bc_inner}) then
translates simply into the condition that
$\tilde\psi^{\rm(in)}(\xi,\eta)$ be continuous at each vertex.
Meanwhile the equation (\ref{eq_inner}) becomes simply
\begin{equation}
  {\rm i}\frac{\partial^2\tilde\psi^{\rm(in)}}
  {\partial\xi\partial\eta}+
  \frac{\partial^4\tilde\psi^{\rm(in)}}{\partial\eta^4}=0.
\label{eq_concatenated}
\end{equation}
This concatenation procedure effectively `irons out' the corners in
the attractor, allowing the solution to proceed continuously.  It is
important to note that the coordinates $(\xi,\eta)$ cover the corner
regions twice.  As noted above, in the corner regions the asymptotic
solution is the sum of the solutions in the segments that meet there.

In fact the boundary condition at vertex $O$ is different because the
perpendicular coordinate $\eta$ undergoes a net focusing by a factor
of $\alpha$ after one loop around the attractor.  The boundary
condition there is
\begin{equation}
  \tilde\psi^{\rm(in)}(\Xi,\alpha\eta)=
  \tilde\psi^{\rm(in)}(0,\eta)+\delta.
\label{bc_concatenated}
\end{equation}
To satisfy this condition requires a kind of self-similar viscous
spreading of the shear layer to compensate for the geometrical
focusing.  Such a self-similar expansion occurs in the solutions of
Moore \& Saffman (1969) for detached shear layers in rotating fluids,
which were used by Rieutord et al. (2001) in their analysis of
inertial waves in a spherical annulus.

Define a similarity variable
\begin{equation}
  \tau=\mu^{-1/3}\eta
\end{equation}
where $\mu=\xi+c$ with $c$ to be determined.  For $\tau$ to map
continuously at the vertex $O$ we require
\begin{equation}
  (\Xi+c)^{-1/3}\alpha\eta=c^{-1/3}\eta
\end{equation}
and therefore
\begin{equation}
  c=\frac{\Xi}{\alpha^3-1}.
\end{equation}
One then writes
\begin{equation}
  \tilde\psi^{\rm(in)}(\xi,\eta)=\frac{\xi}{\Xi}\delta+
  \Psi(\mu,\tau)
\label{sol_inner}
\end{equation}
to accommodate the net increase around the loop.  Here $\mu$ runs from
$c$ to $c+\Xi$ around the attractor.  Equation (\ref{eq_concatenated})
translates into
\begin{equation}
  {\rm i}\left(\frac{\partial}{\partial\mu}-
  \frac{\tau}{3\mu}\frac{\partial}{\partial\tau}\right)
  \left(\mu^{-1/3}\frac{\partial\Psi}{\partial\tau}\right)+
  \mu^{-4/3}\frac{\partial^4\Psi}{\partial\tau^4}=0
\end{equation}
and the boundary condition (\ref{bc_concatenated}) requires simple
continuity:
\begin{equation}
  \Psi(\Xi+c,\tau)=\Psi(c,\tau).
\end{equation}

Making the further transformation
\begin{equation}
  \mu=c\,{\rm e}^\lambda,
\end{equation}
where $\lambda$ runs from $0$ to $3\ln\alpha$, one obtains
\begin{equation}
  {\rm i}\frac{\partial^2\Psi}{\partial\lambda\partial\tau}-
  \frac{{\rm i}}{3}\frac{\partial\Psi}{\partial\tau}-
  \frac{{\rm i}\tau}{3}\frac{\partial^2\Psi}{\partial\tau^2}+
  \frac{\partial^4\Psi}{\partial\tau^4}=0.
\end{equation}
The variable $\lambda$ is now periodic and ignorable, so the solutions
are found by separation of variables to be of the form
\begin{equation}
  \Psi_n={\rm e}^{{\rm i}k_n\lambda}\chi_n(\tau),
\end{equation}
where $k_n=2n\pi/(3\ln\alpha)$, $n\in{\bf Z}$ and $\chi_n$ satisfies the
ordinary differential equation
\begin{equation}
  -\left(k_n+\frac{{\rm i}}{3}\right)\chi'_n-\frac{{\rm i}\tau}{3}\chi''_n+
  \chi''''_n=0.
\end{equation}
Solutions that do not diverge superexponentially as $|\tau|\to\infty$
can be obtained by the Laplace transform method as in Moore \& Saffman
(1969), leading to the integral representation
\begin{equation}
  \chi'_n(\tau)={\rm i}\int_0^\infty{\rm e}^{-{\rm i}p\tau}{\rm e}^{-p^3}
  p^{-3{\rm i}k_n}\,{\rm d}p.
\label{int_rep}
\end{equation}
(These functions are related to Moore--Saffman functions of complex
order and can be expressed in terms of generalized hypergeometric
functions.)  The asymptotic behaviour is
\begin{eqnarray}
  \chi'_n(\tau)&\sim&(-3{\rm i}k_n)!\,{\rm e}^{-(3/2)\pi k_n}
  \tau^{-1+3{\rm i}k_n}
  \qquad\,\,\,\,\hbox{as $\tau\to+\infty$},\label{tauplusinfty}\\
  \chi'_n(\tau)&\sim&-(-3{\rm i}k_n)!\,{\rm e}^{(3/2)\pi k_n}
  |\tau|^{-1+3{\rm i}k_n}
  \qquad\hbox{as $\tau\to-\infty$}.\label{tauminusinfty}
\end{eqnarray}
Therefore $\chi_0(\tau)$ diverges logarithmically as
$\tau\to\pm\infty$, while $\chi_n(\tau)$ for $n\ne0$ is bounded but
oscillates indefinitely.  One may define $\chi_n(\tau)$ uniquely such
that $\chi_n(0)=0$, say.  Then
\begin{equation}
  \chi_n(\tau)\sim\frac{(-3{\rm i}k_n)!}{3{\rm i}k_n}{\rm e}^{\mp(3/2)\pi k_n}
  |\tau|^{3{\rm i}k_n}+c_n^\pm\qquad
  \hbox{as $\tau\to\pm\infty$}
\end{equation}
for some constants $c_n^\pm$.  In the case $n=0$, however,
\begin{equation}
  \chi_0(\tau)\sim\ln|\tau|+c_0^\pm\qquad
  \hbox{as $\tau\to\pm\infty$}.
\end{equation}

The general solution of the inner problem is therefore given by
equation (\ref{sol_inner}) with
\begin{equation}
  \Psi(\mu,\tau)=\sum_{n=-\infty}^\infty\mu^{{\rm i}k_n}
  \left[a_n\chi_n(\tau)+b_n\right],
\end{equation}
where $a_n$ and $b_n$ are undetermined coefficients.  Noting that
$|\tau|^3\mu=|\eta|^3$ one finds the outer limit of the inner solution
to be
\begin{eqnarray}
  \lefteqn{\tilde\psi^{\rm(in)}(\xi,\eta)\sim\frac{\xi}{\Xi}\delta+
  a_0\ln|\eta|-\frac{a_0}{3}\ln(\xi+c)+a_0c_0^\pm+b_0}&\nonumber\\
  &&\qquad+\sum_{n\ne0}\left[a_n\frac{(-3{\rm i}k_n)!}{3{\rm i}k_n}
  {\rm e}^{\mp(3/2)\pi k_n}|\eta|^{3{\rm i}k_n}+
  (a_nc_n^\pm+b_n)(\xi+c)^{{\rm i}k_n}\right]
\label{outer_lim_inner}
\end{eqnarray}
as $\eta\to\pm\infty$.

\subsection{Asymptotic matching}

The aim is now to determine the coefficients $a_n$ by matching the
outer limit of the inner solution to the inner limit of the outer
solution.  (The coefficients $b_n$ will not be required for a
calculation of the asymptotic dissipation rate.)  I consider the
matching on segment~1; it is straightforward to show that the solution
then matches in the same way on the other segments.

The inner limit of the outer solution near segment~1 is, for
$y>0$ or $y<0$ respectively,
\begin{equation}
  \psi^{\rm(out)}(x,y)\sim g(0)-
  \left(-\frac{\ln|y|}{\ln\alpha}\right)\delta-
  H^\pm(\theta^\pm),
\end{equation}
where again
\begin{equation}
  \theta^\pm=\left(-\frac{\ln|y|}{\ln\alpha}\right)\bmod1.
\end{equation}
To compare this solution with the outer limit of the inner solution,
equation (\ref{outer_lim_inner}), note that $\eta=\epsilon^{-1}y$
and therefore
\begin{equation}
  |\eta|^{3{\rm i}k_n}={\rm e}^{2n\pi{\rm i}\ln|\eta|/\ln\alpha}=
  {\rm e}^{-2n\pi{\rm i}\theta^\pm}{\rm e}^{-2n\pi{\rm i}\ln\epsilon/
  \ln\alpha}.
\end{equation}

I consider first the positive side $y>0$ of the attractor and identify
\begin{equation}
  a_0=\frac{\delta}{\ln\alpha},
\label{eq_a0}
\end{equation}
\begin{equation}
  a_n=-\frac{3{\rm i}k_n}{(-3{\rm i}k_n)!}{\rm e}^{(3/2)\pi k_n}
  {\rm e}^{2n\pi{\rm i}\ln\epsilon/\ln\alpha}H^+_{-n},\qquad
  n\ne0.
\end{equation}
Matching on the negative side yields the same expression for $a_0$,
and
\begin{equation}
  a_n=-\frac{3{\rm i}k_n}{(-3{\rm i}k_n)!}{\rm e}^{-(3/2)\pi k_n}
  {\rm e}^{2n\pi{\rm i}\ln\epsilon/\ln\alpha}H^-_{-n},\qquad
  n\ne0.
\end{equation}
It follows that the functions $H^+(\theta^+)$ and $H^-(\theta^-)$ are
related to each other in the Fourier domain by\footnote{The mean
values of $H^\pm$, $H^\pm_0$, are not determined by the matching
procedure.  The relation (\ref{match}) implies that, if $H^\pm$ are
analytic functions, they are identical except that their arguments are
shifted by ${\rm i}\pi/\ln\alpha$ in the complex plane and they also
differ in value by an additive constant.}
\begin{equation}
  H^+_n={\rm e}^{3\pi k_n}H^-_n,\qquad n\ne0.
\label{match}
\end{equation}
Physically, this relation is achieved by a viscous diffusion across
the shear layer, which allows information to be transmitted across the
characteristics of the inviscid equation.

Now using the identity
\begin{equation}
  |({\rm i}x)!|^2=\frac{\pi x}{\sinh(\pi x)}
\end{equation}
for real $x$, one obtains
\begin{equation}
  |a_{-n}|^2=\frac{3k_n}{2\pi}\left(|H^+_n|^2-|H^-_n|^2\right),\qquad
  n\ne0.
\label{eq_amnsq}
\end{equation}

\subsection{The dissipation rate}

The energy dissipation rate $D$ is of order unity in the limit
$\epsilon\to0$ and is dominated by the inner region.  To see this,
consider the following order-of-magnitude estimates, bearing in mind
that the dissipation rate is the area integral of the viscosity times
the square of the vorticity.  In the outer region the streamfunction,
velocity and vorticity are all $O(1)$, the area of the region is
$O(1)$ and the dissipation rate is therefore $O(\epsilon^3)$; it is
small just because the viscosity is small.  In the boundary layers
the velocity is $O(1)$, the vorticity is $O(\epsilon^{-3/2})$, the
area is $O(\epsilon^{3/2})$ and the dissipation rate is therefore
$O(\epsilon^{3/2})$.  Finally, in the inner region the streamfunction
is $O(1)$, the velocity is $O(\epsilon^{-1})$, the vorticity is
$O(\epsilon^{-2})$, the area is $O(\epsilon)$ and the dissipation rate
is therefore $O(1)$.  I now calculate an exact asymptotic expression
for the dissipation rate.

The element of area in segment~$j$ of the inner region is
\begin{equation}
  {\rm d}A={\rm d}x\,{\rm d}y=\epsilon\,{\rm d}\xi_j\,{\rm d}\eta_j=
  \epsilon f_j^{-4}{\rm d}\xi\,{\rm d}\eta.
\end{equation}
Under the transformation from $(\xi,\eta)$ to $(\mu,\tau)$ this
becomes
\begin{equation}
  {\rm d}A=\epsilon f_j^{-4}\mu^{1/3}\,{\rm d}\mu\,{\rm d}\tau.
\end{equation}
The vorticity is dominated by perpendicular derivatives of the
streamfunction:
\begin{equation}
  \nabla^2\psi\sim\epsilon^{-2}
  \frac{\partial^2\tilde\psi^{\rm(in)}}{\partial\eta_j^2}=
  \epsilon^{-2}f_j^2\mu^{-2/3}
  \frac{\partial^2\Psi}{\partial\tau^2}.
\end{equation}
The dissipation rate at leading order therefore simplifies to
\begin{equation}
  D\sim\frac{1}{2}\int_{-\infty}^\infty\int_0^{3\ln\alpha}
  \left|\frac{\partial^2\Psi}{\partial\tau^2}\right|^2
  \,{\rm d}\lambda\,{\rm d}\tau.
\end{equation}
The different `modes' proportional to ${\rm e}^{{\rm i}k_n\lambda}$
add in quadrature with the result
\begin{equation}
  D\sim\frac{3}{2}\ln\alpha\sum_{n=-\infty}^\infty|a_n|^2d_n,
\end{equation}
where
\begin{equation}
  d_n=\int_{-\infty}^\infty\left|\chi''_n(\tau)\right|^2\,{\rm d}\tau
\end{equation}
is just a property of the basis functions.  The integrals $d_n$ are
convergent; the dissipation is well localized even if the
streamfunction is not.  To evaluate $d_n$ I use the integral
representation (\ref{int_rep}) to obtain
\begin{equation}
  d_n=\int_{-\infty}^\infty\int_0^\infty\int_0^\infty
  {\rm e}^{{\rm i}(q-p)\tau}{\rm e}^{-p^3-q^3}
  \left(\frac{q}{p}\right)^{3{\rm i}k_n}
  pq\,{\rm d}p\,{\rm d}q\,{\rm d}\tau.
\end{equation}
The integration with respect to $\tau$ is carried out first, using the
identity
\begin{equation}
  \int_{-\infty}^\infty{\rm e}^{{\rm i}(q-p)\tau}\,{\rm d}\tau=2\pi\delta(q-p),
\end{equation}
and leading to the simple result
\begin{equation}
  d_n=2\pi\int_0^\infty{\rm e}^{-2p^3}p^2\,{\rm d}p=\frac{\pi}{3}.
\end{equation}
The asymptotic dissipation rate is therefore
\begin{equation}
  D\sim\frac{\pi}{2}\ln\alpha\sum_{n=-\infty}^\infty|a_n|^2.
\label{result1}
\end{equation}
Note that $|a_n|^2$ can be related to the outer solution by equation
(\ref{eq_a0}) or (\ref{eq_amnsq}).

\section{A problem with a non-viscous damping force}
\label{s:non-viscous}

\subsection{Motivation}

In this section I investigate a related problem in which the weak
viscous force, which depends on second derivatives of the velocity, is
replaced by a weak damping force of non-viscous form, depending on the
velocity itself.  I show below that the same asymptotic dissipation
rate is obtained in this case, implying that the dissipation rate is
robust, being independent of both the magnitude and the form of the
damping force in the limit that the damping is weak.

One reason for investigating this robustness is that the true
mechanism by which tidal disturbances in astrophysical bodies, such as
inertial waves in the convective regions of giant planets, are damped
is uncertain.  The true viscosity is exceedingly small, and the waves
may decay more readily by other means.  Their interaction with
convective motions might be modelled in terms of an eddy viscosity,
but the validity of such an approach is unclear and estimates of the
relevant effective viscosity vary widely.  Alternatively, inertial
waves may undergo Ohmic damping in the presence of a magnetic field
since they couple to Alfv\'en waves on small scales.  They may also
experience nonlinear parametric decay into waves of shorter
wavelength.  The last two mechanisms are probably modelled more
accurately, although still only crudely, by using a `frictional'
damping force proportional to the velocity, rather than a viscous
force.

Another way to view the following analysis is that it results from the
Landau prescription in which the frequency of the forced wave is given
a small positive imaginary part, thereby rendering the inviscid
problem soluble, and a limit is taken in which the imaginary part of
the frequency tends to zero.  Such an approach might be justified by a
consideration of the late-time asymptotic solution of the inviscid
initial-value problem constructed by a Laplace transform method.

\subsection{Sketch of the analysis}

Instead of equation (\ref{eq_del4}) I now consider
\begin{equation}
  {\rm i}\frac{\partial^2\psi}{\partial x\partial y}-\epsilon\nabla^2\psi=f,
\label{eq_del2}
\end{equation}
with the boundary condition
\begin{equation}
  \psi=0\qquad\hbox{on $\partial{\cal D}$}.
\end{equation}
The dissipation rate for this problem is
\begin{equation}
  D=\frac{1}{2}\int_{\cal D}\epsilon|\bnabla\psi|^2\,{\rm d}A=
  \frac{1}{2}{\rm Re}\int_{\cal D}\psi^*f\,{\rm d}A.
\end{equation}

The analysis proceeds in a very similar way to the previous
calculation and it is necessary only to point out the essential
differences.  One simplification is that no viscous boundary layers
occur, although in fact it was not necessary to consider the boundary
layers of the previous problem in any detail.  The essential
differences appear in the analysis of the inner region, which is again
of width $O(\epsilon)$.  The reduced equation (\ref{eq_inner}) in
segment~$j$ becomes
\begin{equation}
  {\rm i}\frac{\partial^2\tilde\psi_j^{\rm(in)}}
  {\partial\xi_j\partial\eta_j}-
  \frac{\partial^2\tilde\psi_j^{\rm(in)}}{\partial\eta_j^2}=0.
\label{eq_inner_del2}
\end{equation}
The factors of $f^3$ and $t^3$ in equations (\ref{xi}--\ref{Xi})
should be replaced by single powers and the concatenated inner
equation (\ref{eq_concatenated}) becomes
\begin{equation}
  {\rm i}\frac{\partial^2\tilde\psi^{\rm(in)}}{\partial\xi\partial\eta}-
  \frac{\partial^2\tilde\psi^{\rm(in)}}{\partial\eta^2}=0.
\end{equation}
A similarity solution satisfying the boundary conditions can be
obtained in an analogous way, now using variables $\tau=\eta/(\xi+c)$
and $\lambda=\ln(\xi/c+1)$ with $c=\Xi/(\alpha-1)$.  The parallel
coordinate $\lambda$ runs from $0$ to $\ln\alpha$, not $3\ln\alpha$ as
previously.  The inner equation in similarity variables now reads
\begin{equation}
  {\rm i}\frac{\partial^2\Psi}{\partial\lambda\partial\tau}-
  {\rm i}\frac{\partial\Psi}{\partial\tau}-
  {\rm i}\tau\frac{\partial^2\Psi}{\partial\tau^2}-
  \frac{\partial^2\Psi}{\partial\tau^2}=0
\end{equation}
and has solutions
\begin{equation}
  \Psi_n={\rm e}^{3{\rm i}k_n\lambda}\chi_n(\tau)
\end{equation}
where I define $k_n=2n\pi/(3\ln\alpha)$ as previously, and $\chi_n$
now satisfies
\begin{equation}
  -(3k_n+{\rm i})\chi'_n-{\rm i}\tau\chi''_n-\chi''_n=0.
\end{equation}
The Laplace transform method leads to
\begin{equation}
  \chi'_n(\tau)={\rm i}\int_0^\infty{\rm e}^{-{\rm i}p\tau}{\rm e}^{-p}
  p^{-3{\rm i}k_n}\,{\rm d}p=
  (-3{\rm i}k_n)!\,{\rm e}^{-(3/2)\pi k_n}(\tau-{\rm i})^{-1+3{\rm i}k_n},
\end{equation}
where the branch cut is confined to the upper half-plane.  The
asymptotic behaviour as $\tau\to\pm\infty$ is precisely as in
equations (\ref{tauplusinfty}) and (\ref{tauminusinfty}), and the
asymptotic matching therefore yields precisely the same values for the
coefficients $a_n$.

Finally, the asymptotic dissipation rate is computed as
\begin{equation}
  D\sim\frac{1}{2}\int_{-\infty}^\infty\int_0^{\ln\alpha}
  \left|\frac{\partial\Psi}{\partial\tau}\right|^2
  \,{\rm d}\lambda\,{\rm d}\tau
  \sim\frac{\ln\alpha}{2}\sum_{n=-\infty}^\infty|a_n|^2d_n,
\end{equation}
where now
\begin{equation}
  d_n=\int_{-\infty}^\infty|\chi'_n(\tau)|^2\,{\rm d}\tau=\pi.
\end{equation}
The end result is that the asymptotic dissipation rate is exactly the
same as for the viscous problem,
\begin{equation}
  D\sim\frac{\pi}{2}\ln\alpha\sum_{n=-\infty}^\infty|a_n|^2.
\end{equation}

\section{Numerical solutions}
\label{s:numerical}

Numerical solutions of equations (\ref{eq_maas}), (\ref{eq_del4}) and
(\ref{eq_del2}) in various quadrilateral domains have been obtained by
the following method.  A nonlinear algebraic coordinate transformation
is first applied to map the quadrilateral domain on to the unit
square.  The equations are then discretized using centred second-order
finite differences, resulting in a large linear system in block
tridiagonal form.  This is solved by a direct method and the total
dissipation rate is evaluated, again to second-order accuracy, from a
numerical integration based on either equation (\ref{d1}) or equation
(\ref{d2}) (or their equivalents for the other differential
equations).  Typically equation (\ref{d2}) gives a much more accurate
numerical value for $D$, as it does not require a numerical
differentiation of the solution.

\begin{figure}
  \centerline{\epsfysize10cm\epsfbox{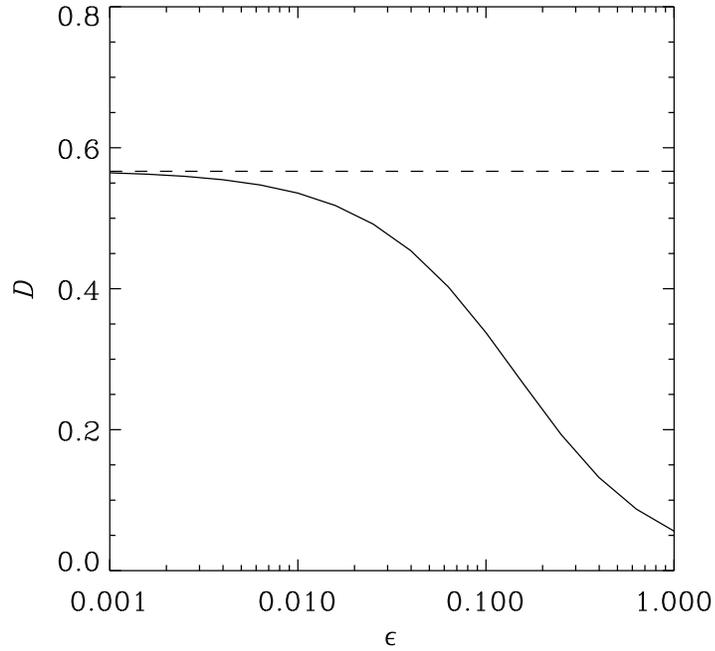}}
  \caption{Dissipation rate versus damping parameter for equation
  (\ref{eq_del2}) in a tilted square domain ($X=Y=1$, $t_i=2$) with
  uniform forcing $(f=1)$.  The dashed line indicates the limiting
  value expected on the basis of the asymptotic analysis.}
\label{f:dissipation}
\end{figure}

One of the simplest problems involving a wave attractor is to solve
equation (\ref{eq_del2}) in a tilted square domain ($X=Y=1$,
$t_1=t_2=t_3=t_4\ne1$) with uniform forcing $(f=1)$.  The dissipation
rate for the case $t_i=2$ is plotted as a function of $\epsilon$ in
figure~\ref{f:dissipation}, showing the anticipated convergence to a
non-zero limit as $\epsilon\to0$.  Of course, a higher numerical
resolution is required when $\epsilon$ is reduced.
Figure~\ref{f:images} shows the spatial distribution of the
dissipation rate for $\epsilon=0.01$ and $0.003$ as calculated with a
numerical resolution of $1000^2$.  As expected from the asymptotic
solution, the dissipation is localized in the neighbourhood of the
attractor and the width of the beam is proportional to $\epsilon$.

\begin{figure}
  \bigskip
%  \centerline{\epsfysize7cm\epsfbox{plot_1e-2_1000.eps}\qquad
%  \epsfysize7cm\epsfbox{plot_3e-3_1000.eps}}
  \vbox to7cm{\vfill\centerline{\Large\it These colour figures are supplied as separate GIF files}\vfill}
  \medskip
  \caption{Spatial distribution of the dissipation rate for the
  problem referred to in figure~\ref{f:dissipation}.  The left and
  right panels are for $\epsilon=0.01$ and $0.003$, respectively.  In
  each case a logarithmic colour scale over three orders of magnitude
  is used.}
\label{f:images}
\end{figure}

\begin{figure}
  \centerline{\epsfysize14cm\epsfbox{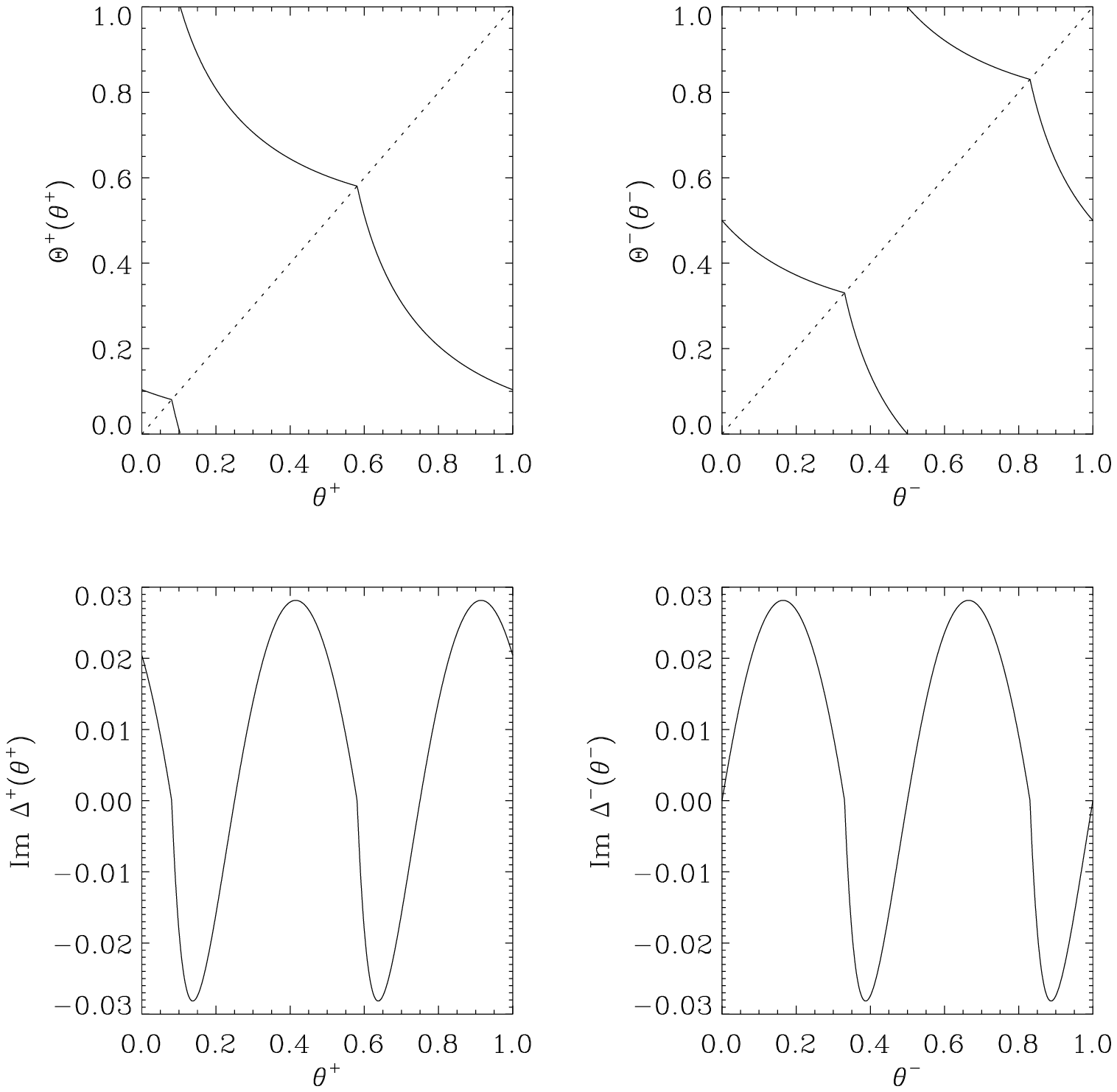}}
  \caption{The ray-mapping and forcing functions in the tilted square
  domain ($X=Y=1$, $t_i=2$) with uniform forcing ($f=1$).}
\label{f:rays}
\end{figure}

To determine the expected limiting value of $D$ from the asymptotic
analytical solution, some calculation is required.  The focusing and
forcing constants $\alpha$ and $\delta$ are easily evaluated.  A
simple programme is then used to follow the propagation of rays in the
quadrilateral domain and thereby to evaluate the ray-mapping and
forcing functions $\Theta^\pm(\theta^\pm)$ and
$\Delta^\pm(\theta^\pm)$; these are shown in figure~\ref{f:rays}.

Next the functional equations
\begin{equation}
  H^\pm(\Theta^\pm(\theta^\pm))-H^\pm(\theta^+)=\Delta^\pm(\theta^\pm)
\label{hpm-delta}
\end{equation}
must be solved, subject to the coupling condition on the Fourier
coefficients
\begin{equation}
  H_n^+={\rm e}^{3\pi k_n}H_n^-,\qquad n\ne0.
\label{hp-hm}
\end{equation}
Equations (\ref{hpm-delta}) can be represented in the Fourier domain
in the form
\begin{equation}
  \sum_{n=-\infty}^\infty M_{mn}^\pm H_n^\pm-H_m^\pm=\Delta_m^\pm,
\label{hpm-delta-fourier}
\end{equation}
where
\begin{equation}
  M_{mn}^\pm=\int_0^1{\rm e}^{-2m\pi{\rm i}\theta^\pm}
  {\rm e}^{2n\pi{\rm i}\Theta^\pm(\theta^\pm)}\,{\rm d}\theta^\pm,
\end{equation}
\begin{equation}
  \Delta_m^\pm=\int_0^1{\rm e}^{-2m\pi{\rm i}\theta^\pm}
  \Delta^\pm(\theta^\pm)\,{\rm d}\theta^\pm.
\end{equation}
Equations (\ref{hpm-delta-fourier}) are semi-redundant because the
symmetry properties $\Theta^\pm(\Theta^\pm(\theta^\pm))=\theta^\pm$
and $\Delta^\pm(\Theta^\pm(\theta^\pm))=-\Delta^\pm(\theta^\pm)$
ensure that the equations merely change sign when acted upon by the
appropriate operator with matrix coefficients $M_{mn}^\pm$.  It is
most convenient for numerical purposes to regard the unknown
quantities as the `large' Fourier coefficients $H_n^+$ ($1\le n\le N$)
and $H_n^-$ ($-N\le n\le-1$) up to some finite truncation order $N$,
substituting for the `small' coefficients using equation
(\ref{hp-hm}).  The property $M_{m0}^\pm=\delta_{m0}$ means that the
quantities $H_0^\pm$ cannot be determined and in any case are not
required to calculate the asymptotic dissipation rate.  In view of the
semi-redundancy it is sufficient to solve only half of the truncated
system of equations (\ref{hpm-delta-fourier}), e.g. the `$+$'
equations for $m>0$ and the `$-$' equations for $m<0$.  The
coefficients $M_{mn}^\pm$ and $\Delta_m^\pm$ are readily calculated to
high accuracy from the computed functions $\Theta^\pm(\theta^\pm)$ and
$\Delta^\pm(\theta^\pm)$.

In practice it is found that, for a smooth forcing function $f$, the
coefficients $a_n$ typically decay rapidly with $n$ so that $a_0$
makes by far the dominant contribution to $D$.  This property is very
convenient because $a_0$ is easily calculated and does not depend on
the details of the global ray mapping.  The dashed line in
figure~\ref{f:dissipation} indicates the expected asymptotic value of
$D$, to which terms other than $a_0$ contribute less than one per
cent.  Convergence to the same dissipation rate is found for the
viscous problem defined by equation (\ref{eq_del4}).

\begin{figure}
  \centerline{\epsfysize12cm\epsfbox{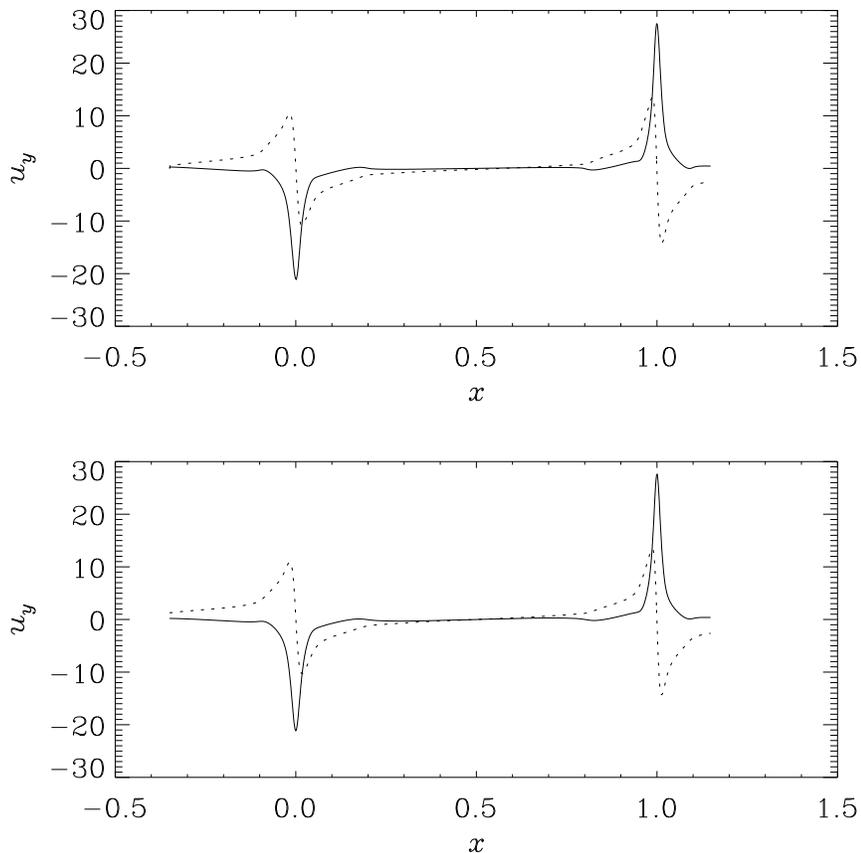}}
  \caption{Cross-section at height $y=0.3$ of the $y$-component of
  velocity, defined by $u_y=-\partial\psi/\partial x$, for the problem
  referred to in figure~\ref{f:dissipation} with $\epsilon=0.01$.  The
  upper panel shows the direct numerical solution, while the lower
  panel shows the sum of the asymptotic inner solutions for segments~2
  and~4.  The real and imaginary parts of $u_y$ are given by the solid
  and dotted lines, respectively.}
\label{f:velocity}
\end{figure}

The asymptotic analytical solution is compared in detail with the
direct numerical solution in Figure~\ref{f:velocity}, which shows the
$y$-component of the velocity along a cross-section through the tilted
square domain at $y=0.3$.  In drawing the asymptotic solution along
this section, the inner solutions corresponding to segments~2 and~4
are summed.  No attempt is made to patch with the outer solution,
which is much smaller and more difficult to determine properly.  Even
with this restriction the agreement is very good at $\epsilon=0.01$.

\section{Discussion}
\label{s:discussion}

In this paper I have considered a prototypical forced linear wave
equation featuring a wave attractor and weak viscous damping.  This
mathematical problem describes the long-term linear response of a
typical bounded fluid to a periodic body force with a frequency within
the range of inertial and gravity waves.  Through asymptotic analysis
confirmed by direct numerical calculations in an illustrative
quadrilateral domain, I have shown that the forced disturbance is
localized in the neighbourhood of the attractor and that the total
dissipation rate is asymptotically independent of the viscosity.  By
considering a related equation with a non-viscous damping force, I
have argued further that the asymptotic dissipation rate is
independent of the nature, as well as the magnitude, of the weak
damping mechanism.

These findings have important consequences for physical problems such
as that considered in \S\ref{s:maas} and motivated by the experiments
of Maas et al. (1997), in which a periodic body force excites internal
gravity waves in a narrow tank of non-rectangular cross-section.  In
that problem the wave attractors exist in certain intervals of
frequency that depend only on the geometry of the container.  The
long-term linear response of the fluid should give rise to a
dissipation rate that is asymptotically independent of the viscosity
and is a smoothly varying function of the forcing frequency within the
interval associated with a given attractor.  This differs markedly
from the case of an upright rectangular tank, in which the response of
the fluid is governed by resonances with the regular inviscid normal
modes that can exist in that container, and the dissipation rate will
be very small except where it is strongly amplified in the vicinity of
the resonant frequencies of the lowest-order modes.  As noted by Maas
et al. (1997), wave attractors have a `finite bandwidth' while the
width of a normal-mode resonance is limited by viscosity.

Ogilvie \& Lin (2004) recently studied the linear response of a giant
planet with an extended convective region to periodic tidal forcing.
Inertial waves propagating in the annular convective region do not
form regular eigenfunctions in the absence of viscosity.  Numerical
solutions indicated that, when the Ekman number is small, the
disturbance tends to be localized on a web of rays.  For intervals of
frequency in which simple wave attractors exist, it was possible to
identify their appearance in the solutions.  In figure~10 of Ogilvie
\& Lin (2004) it can be seen that the dissipation rate appears to have
converged to a value independent of the viscosity in the intervals
where the dissipation rate is largest.  It can be confirmed by tracing
rays in the annulus that these intervals are those associated with the
simplest and most powerfully focusing wave attractors.  Although more
complicated attractors exist outside these intervals, their focusing
power is evidently not great enough to have led to a convergence of
the dissipation rate at an Ekman number of $10^{-8}$.  These numerical
results are qualitatively consistent with the conclusions of the
present paper, and in future work we will attempt to evaluate the
asymptotic dissipation rate for tidal forcing in a spherical annulus
directly using the methods described here.

Forced inertial waves in a rotating incompressible fluid contained in
a spherical annulus were computed previously by Tilgner (1999), who
noted that wave attractors can be detected in the solutions at small
Ekman numbers but that other complicating features are also present.
In his problem, which derives from earlier experimental studies, the
forcing results from a sinusoidal modulation of the rotation rate of
the boundaries, which is communicated to the fluid through viscous
boundary layers.  As the viscosity is reduced, the energy of the
forced disturbance is also diminished.  This behaviour is different
from that found in the present paper with a distributed body force
that is independent of viscosity, such as that resulting from tidal
forcing.

The apparent robustness of the dissipation rate is reassuring when it
is considered that the true mechanism by which the inertial waves are
damped is uncertain and difficult to model accurately.  It is likely
that the result is reasonably robust in other ways as well.  For
example, small corrugations of the boundary may not be so important
for inertial waves as for acoustic waves, which undergo specular
reflection.  Inertial waves of a given frequency can propagate only in
four different directions, and the angle of reflection will not be
changed by a small corrugation.  The wave attractors depend more on
the gross shape of the container than on the precise contours of the
boundary.

It is also noteworthy that the `inner' equations (\ref{eq_inner}) and
(\ref{eq_inner_del2}) studied in this paper are generic in the
following sense.  Provided that the inviscid wave equation is a
hyperbolic second-order linear partial differential equation, its
characteristics can be determined and any wave attractors identified.
Curvilinear coordinates can then be introduced parallel and
perpendicular to each segment of the attractor.  If the equation
contains a weak damping term resulting from a positive operator of
either fourth or second order, then the coordinates can always be
rescaled to obtain equation (\ref{eq_inner}) or equation
(\ref{eq_inner_del2}), respectively.  This is true even if the
inviscid equation has non-constant coefficients and the damping
operator is anisotropic and depends on position, provided that there
are no singularities such as corotation resonances or critical
latitudes.

Further work is required to understand the transition from normal
modes to wave attractors as the geometry of the container is varied.
Ogilvie \& Lin (2004) found that some `memory' of the normal modes of
inertial waves in a full sphere was retained when a small solid core
was introduced.  Even though no inviscid normal modes are believed to
exist in a spherical annulus, the dissipation rate is still enhanced
in the vicinity of the eigenfrequencies of the lowest-order modes of
the full sphere.  An imperfect resonance is possible with such modes,
which have a length-scale larger than the radius of the core and are
not greatly affected by it.  At the same time, introduction of the
solid core allows a much richer response at other frequencies,
associated with wave attractors.  It is possible that in such problems
the attractors achieve complete dominance only at Ekman numbers yet
smaller than those achieved in recent numerical studies.

A remarkable theory of axisymmetric inertial waves in an
incompressible rotating fluid contained in a spherical annulus was
developed by Rieutord et al. (2002), who considered normal modes that
are regularized and damped by viscosity.  Their problem is similar to
that considered in the present paper, but there are also important
differences.  Neglecting some of the effects of curvature so as to
make the equations somewhat simpler, they found asymptotic solutions
for normal modes in the limit of small Ekman number, in excellent
agreement with their numerical calculations.  The modes are localized
near a wave attractor and satisfy an eigenvalue problem analogous to
that for the wavefunction of a quantum-mechanical harmonic oscillator.
The width of the shear layers in these modes scales with $\nu^{1/4}$,
although the authors noted that $\nu^{1/3}$ layers also play a role
when the problem is studied in full spherical geometry.  In contrast,
the solutions described in \S\ref{s:prototypical} have only a
$\nu^{1/3}$ shear layer and satisfy a different inner equation that
has no strictly localized solutions but is forced from the outside.
The reason for the different scalings is that the normal modes
considered by Rieutord et al. (2002) have frequencies that are
asymptotically close to the edge of the band within which the
attractor exists, so that the attractor is very weak and its focusing
power $\alpha$ is asymptotically close to~$1$.  In contrast, the
forced solutions in the present paper are constructed for the more
interesting frequencies that lie properly within the bandwidth of the
attractor, so that $\alpha$ is not very close to~$1$.  It is likely
that a different asymptotic theory of forced oscillations could be
constructed for values of $\alpha$ very close to~$1$, which would more
closely resemble the analysis of Rieutord et al. (2002).  However, it
still remains to be demonstrated in detail how the analysis in the
present paper applies to forced oscillations in spherical geometry.

\begin{acknowledgments}
  I thank Doug Lin for introducing me to this problem and for ongoing
  discussions, as well as for hospitality at UC Santa Cruz where some
  of this work was carried out.  I also thank Rainer Hollerbach, John
  Papaloizou, Michel Rieutord and Yanqin Wu for helpful discussions,
  and the anonymous referees for their useful suggestions.  I
  acknowledge the support of the Royal Society through a University
  Research Fellowship.
\end{acknowledgments}

\appendix

\section{Character of the inviscid linearized equations}
\label{a:character}

The equations governing the dynamics of an ideal fluid with negligible
self-gravity may be written
\begin{equation}
  \frac{{\rm D}\bu}{{\rm D}t}=-\frac{1}{\rho}\bnabla p-\bnabla\Phi,
\end{equation}
\begin{equation}
  \frac{{\rm D}\rho}{{\rm D}t}=-\rho\bnabla\bcdot\bu,
\end{equation}
\begin{equation}
  \frac{{\rm D}p}{{\rm D}t}=-\gamma p\bnabla\bcdot\bu,
\end{equation}
where $\Phi$ is the external gravitational potential and the notation
is standard.  Let $(r,\phi,z)$ be cylindrical polar coordinates and
consider a basic state consisting of an axisymmetric fluid body with
angular velocity $\Omega(r,z)$ and no meridional flow.  For small
perturbations of the form
\begin{equation}
  {\rm Re}\left[\bu'(r,z)\,{\rm e}^{-{\rm i}\omega t+{\rm i}m\phi}\right],
\end{equation}
etc., the linearized equations read
\begin{equation}
  -{\rm i}\hat\omega u_r'-2\Omega u_\phi'=
  -\frac{1}{\rho}\frac{\partial p'}{\partial r}+
  \frac{\rho'}{\rho^2}\frac{\partial p}{\partial r},
\end{equation}
\begin{equation}
  -{\rm i}\hat\omega u_\phi'+\frac{1}{r}
  \left(u_r'\frac{\partial}{\partial r}+u_z'\frac{\partial}{\partial z}\right)
  (r^2\Omega)=-\frac{{\rm i}mp'}{\rho r},
\end{equation}
\begin{equation}
  -{\rm i}\hat\omega u_z'=-\frac{1}{\rho}\frac{\partial p'}{\partial z}+
  \frac{\rho'}{\rho^2}\frac{\partial p}{\partial z},
\end{equation}
\begin{equation}
  -{\rm i}\hat\omega\rho'+u_r'\frac{\partial\rho}{\partial r}+
  u_z'\frac{\partial\rho}{\partial z}=
  -\rho\left[\frac{1}{r}\frac{\partial}{\partial r}(ru_r')+
  \frac{{\rm i}mu_\phi'}{r}+\frac{\partial u_z'}{\partial z}\right],
\end{equation}
\begin{equation}
  -{\rm i}\hat\omega p'+u_r'\frac{\partial p}{\partial r}+
  u_z'\frac{\partial p}{\partial z}=
  -\gamma p\left[\frac{1}{r}\frac{\partial}{\partial r}(ru_r')+
  \frac{{\rm i}mu_\phi'}{r}+\frac{\partial u_z'}{\partial z}\right],
\end{equation}
where $\hat\omega=\omega-m\Omega$ is the Doppler-shifted wave
frequency.  Eliminate $u_\phi'$ and $\rho'$ to obtain
\begin{equation}
  (\hat\omega^2-A)u_r'-Bu_z'=
  -\frac{{\rm i}\hat\omega}{\rho}\left(\frac{\partial p'}{\partial r}-
  \frac{\partial p}{\partial r}\frac{p'}{\gamma p}\right)+
  2\Omega\frac{{\rm i}mp'}{\rho r},
\end{equation}
\begin{equation}
  -Cu_r'+(\hat\omega^2-D)u_z'=
  -\frac{{\rm i}\hat\omega}{\rho}\left(\frac{\partial p'}{\partial z}-
  \frac{\partial p}{\partial z}\frac{p'}{\gamma p}\right),
\end{equation}
where
\begin{equation}
  A=\frac{2\Omega}{r}\frac{\partial}{\partial r}(r^2\Omega)-
  \frac{1}{\rho}\frac{\partial p}{\partial r}
  \left(\frac{1}{\gamma p}\frac{\partial p}{\partial r}-
  \frac{1}{\rho}\frac{\partial\rho}{\partial r}\right),
\end{equation}
\begin{equation}
  B=\frac{2\Omega}{r}\frac{\partial}{\partial z}(r^2\Omega)-
  \frac{1}{\rho}\frac{\partial p}{\partial r}
  \left(\frac{1}{\gamma p}\frac{\partial p}{\partial z}-
  \frac{1}{\rho}\frac{\partial\rho}{\partial z}\right),
\end{equation}
\begin{equation}
  C=-\frac{1}{\rho}\frac{\partial p}{\partial z}
  \left(\frac{1}{\gamma p}\frac{\partial p}{\partial r}-
  \frac{1}{\rho}\frac{\partial\rho}{\partial r}\right),
\end{equation}
\begin{equation}
  D=-\frac{1}{\rho}\frac{\partial p}{\partial z}
  \left(\frac{1}{\gamma p}\frac{\partial p}{\partial z}-
  \frac{1}{\rho}\frac{\partial\rho}{\partial z}\right).
\end{equation}
Then eliminate $u_r'$ and $u_z'$ to obtain
\begin{equation}
  (\hat\omega^2-D)\frac{\partial^2p'}{\partial r^2}+
  (B+C)\frac{\partial^2p'}{\partial r\partial z}+
  (\hat\omega^2-A)\frac{\partial^2p'}{\partial z^2}+
  \hbox{(terms in $p'$ and $\bnabla p'$)}=0.
\end{equation}
This equation is hyperbolic when
\begin{equation}
  4(\hat\omega^2-A)(\hat\omega^2-D)<(B+C)^2,
\end{equation}
i.e. for frequencies such that\footnote{The H\o iland criteria for
stability with respect to axisymmetric perturbations are $A+D>0$ and
$AD>BC$ (e.g. Tassoul 1978).  For a H\o iland-stable basic state the
hyperbolic range of frequencies may or may not include
$\hat\omega=0$.}
\begin{equation}
  (A+D)^2-\left[(B+C)^2+(A-D)^2\right]^{1/2}<2\hat\omega^2<
  (A+D)^2+\left[(B+C)^2+(A-D)^2\right]^{1/2}.
\end{equation}

\section{Vortical effective forcing}
\label{a:vortical}

Ogilvie \& Lin (2004) studied the linear response of a rotating giant
planet (or star) to tidal forcing in circumstances in which the
forcing frequency is comparable to the angular velocity $\Omega$ of
the body but small compared to its dynamical frequency
$(GM/R^3)^{1/2}$.  They used a consistent asymptotic expansion to
simplify the equations satisfied by the tidal disturbance in both
convective and radiative regions.  The tidal force, which derives from
a perturbing gravitational potential, is taken up by a
quasi-hydrostatic adjustment (a tidal bulge) known as the equilibrium
tide.  Because the tidal frequency is non-zero, the motion of the
bulge gives rise to a non-zero velocity field $\bu_{\rm e}$ associated
with the equilibrium tide, which can be derived analytically.  The
existence of inertial forces associated with this flow mean that the
quasi-hydrostatic bulge does not quite satisfy the equation of motion
and a residual force appears.

In quantitative terms, the total velocity perturbation $\bu$ in an
adiabatically stratified (i.e. convective) region satisfies the
equations
\begin{equation}
  -{\rm i}\hat\omega\bu+2{\bf\Omega}\times\bu=-\bnabla W,
\end{equation}
\begin{equation}
  \frac{\partial\rho'_{\rm e}}{\partial t}+\bnabla\bcdot(\rho\bu)=0,
\end{equation}
where $\hat\omega=\omega-m\Omega$ is the Doppler-shifted frequency as
in Appendix~\ref{a:character}, $W$ is the pressure perturbation
divided by the density, plus the total gravitational potential
perturbation, and $\rho'_{\rm e}$ is the density perturbation
associated with the equilibrium tide (any additional density
perturbation being much smaller).  Here it is assumed for simplicity
that the fluid is uniformly rotating and the viscosity is neglected.
Now since
\begin{equation}
  \frac{\partial\rho'_{\rm e}}{\partial t}+\bnabla\bcdot(\rho\bu_{\rm e})=0,
\end{equation}
the residual `dynamical tide' $\bu_{\rm d}=\bu-\bu_{\rm e}$ is found
to satisfy the equations
\begin{equation}
  -{\rm i}\hat\omega\bu_{\rm d}+2{\bf\Omega}\times\bu_{\rm d}=-\bnabla W+\ba,
\end{equation}
\begin{equation}
  \bnabla\bcdot(\rho\bu_{\rm d})=0,
\end{equation}
where
\begin{equation}
  \ba={\rm i}\hat\omega\bu_{\rm e}-2{\bf\Omega}\times\bu_{\rm e}.
\end{equation}
The dynamical tide therefore obeys the anelastic approximation, in
which the modified pressure perturbation $W$ must adjust to satisfy
the constraint $\bnabla\bcdot(\rho\bu_{\rm d})=0$, and is driven by an
effective body force $\ba$ per unit mass.  In general this effective
force is vortical, because $\bnabla\times\ba\ne{\bf0}$.


\begin{thebibliography}{}
 
\bibitem[Aldridge \& Lumb (1980)]{AL87}
  \textsc{Aldridge, K. D. \& Lumb, L. I.} 1987
  Inertial waves identified in the Earth's fluid outer core.
  \textit{Nature} \textbf{325}, 421--423.

\bibitem[Bretherton (1964)]{B64}
  \textsc{Bretherton, F. P.} 1964
  Low-frequency oscillations trapped near the equator.
  \textit{Tellus} \textbf{16}, 181--185.

\bibitem[Chandrasekhar (1961)]{C61}
  \textsc{Chandrasekhar, S.} 1961
  \textit{Hydrodynamic and Hydromagnetic Stability.}  Oxford University Press.

\bibitem[Christensen-Dalsgaard (2002)]{C02}
  \textsc{Christensen-Dalsgaard, J.} 2002
  Helioseismology.
  \textit{Rev. Mod. Phys.} \textbf{74}, 1073--1129.

\bibitem[Goldreich \& Tremaine (1980)]{GT80}
  \textsc{Goldreich, P. \& Tremaine, S.} 1980
  Disk--satellite interactions.
  \textit{Astrophys. J.} \textbf{241}, 425--441.

\bibitem[Greenspan (1968)]{G68}
  \textsc{Greenspan, H. P.} 1968
  \textit{The Theory of Rotating Fluids.}  Cambridge University Press.

\bibitem[Hollerbach \& Kerswell (1995)]{HK95}
  \textsc{Hollerbach, R. \& Kerswell, R. R.} 1995
  Oscillatory internal shear layers in rotating and precessing flows.
  \textit{J. Fluid Mech.} \textbf{298}, 327--339.
 
\bibitem[Ioannou \& Lindzen (1993)]{IL93}
  \textsc{Ioannou, P. J. \& Lindzen, R. S.} 1993
  Gravitational tides in the outer planets.  I. Implications of classical
  tidal theory.
  \textit{Astrophys. J.} \textbf{406}, 252--265.
 
\bibitem[Kato (2001)]{K01}
  \textsc{Kato, S.} 2001
  Basic properties of thin-disk oscillations.
  \textit{Proc. Astron. Soc. Japan} \textbf{53}, 1--24.
 
\bibitem[Maas (2001)]{M01}
  \textsc{Maas, L. R. M.} 2001
  Wave focusing and ensuing mean flow due to symmetry breaking in rotating
  fluids.
  \textit{J. Fluid Mech.} \textbf{437}, 13--28.

\bibitem[Maas \& Lam (1995)]{ML95}
  \textsc{Maas, L. R. M. \& Lam, F.-P. A.} 1995
  Geometric focusing of internal waves.
  \textit{J. Fluid Mech.} \textbf{300}, 1--41.

\bibitem[Maas et al. (1997)]{MBSL97}
  \textsc{Maas, L. R. M., Benielli, D., Sommeria, J. \& Lam, F.-P. A.} 1997
  Observation of an internal wave attractor in a confined, stably stratified
  fluid.
  \textit{Nature} \textbf{388}, 557--561.
 
\bibitem[Manders \& Maas (2003)]{MM03}
  \textsc{Manders, A. M. M. \& Maas, L. R. M.} 2003
  Observations of inertial waves in a rectangular basin with one sloping
  boundary.
  \textit{J. Fluid Mech.} \textbf{493}, 59--88.

\bibitem[Moore \& Saffman (1969)]{MS69}
  \textsc{Moore, D. W. \& Saffman, P. G.} 1969
  The structure of free vertical shear layers in a rotating fluid and the
  motion produced by a slowly rising body.
  \textit{Phil. Trans. R. Soc. London A} \textbf{264}, 597--634.

\bibitem[Ogilvie \& Lin (2004)]{OL04}
  \textsc{Ogilvie, G. I. \& Lin, D. N. C.} 2004
  Tidal dissipation in rotating giant planets.
  \textit{Astrophys. J.} \textbf{610}, 477--509.

\bibitem[Rieutord, Georgeot \& Valdettaro (2001)]{RGV01}
  \textsc{Rieutord, M., Georgeot, B. \& Valdettaro, L.} 2001
  Inertial waves in a rotating spherical shell: attractors and asymptotic
  spectrum.
  \textit{J. Fluid Mech.} \textbf{435}, 103--144.

\bibitem[Rieutord \& Valdettaro (1997)]{RV97}
  \textsc{Rieutord, M. \& Valdettaro, L.} 1997
  Inertial waves in a rotating spherical shell.
  \textit{J. Fluid Mech.} \textbf{341}, 77--99.

\bibitem[Rieutord, Valdettaro \& Georgeot (2002)]{RVG02}
  \textsc{Rieutord, M., Valdettaro, L. \& Georgeot, B.} 2002
   Analysis of singular inertial modes in a spherical shell: the slender
   toroidal shell model.
  \textit{J. Fluid Mech.} \textbf{463}, 345--360.

\bibitem[Savonije \& Papaloizou (1997)]{SP97}
  \textsc{Savonije, G. J. \& Papaloizou, J. C. B.} 1997
  Non-adiabatic tidal forcing of a massive, uniformly rotating star -- II.
  The low-frequency, inertial regime.
  \textit{Mon. Not. R. Astron. Soc.} \textbf{291}, 633--650.

\bibitem[Savonije \& Witte (2002)]{WS02}
  \textsc{Savonije, G. J. \& Witte, M. G.} 2002
  Tidal interaction of a rotating 1 $M_\odot$ star with a binary companion.
  \textit{Astron. Astrophys.} \textbf{386}, 211--221.
 
\bibitem[Savonije, Papaloizou \& Alberts (1995)]{SPA95}
  \textsc{Savonije, G. J., Papaloizou, J. C. B. \& Alberts, F.} 1995
  Non-adiabatic tidal forcing of a massive, uniformly rotating star.
  \textit{Mon. Not. R. Astron. Soc.} \textbf{277}, 471--496.

\bibitem[Stern (1963)]{S63}
  \textsc{Stern, M. E.} 1963
  Trapping of low-frequency oscillations in an equatorial `boundary layer'.
  \textit{Tellus} \textbf{15}, 246--250.

\bibitem[Stewartson (1972)]{S72}
  \textsc{Stewartson, K.} 1972
  On trapped oscillations in a slightly viscous rotating fluid.
  \textit{J. Fluid Mech.} \textbf{54}, 749--761.

\bibitem[Stewartson \& Rickard (1969)]{SR69}
  \textsc{Stewartson, K. \& Rickard, J. A.} 1969
  Pathological oscillations of a rotating fluid.
  \textit{J. Fluid Mech.} \textbf{35}, 759--773.

\bibitem[Tassoul (1978)]{T78}
  \textsc{Tassoul, J.-L.} 1978
  \textit{Theory of Rotating Stars.}  Princeton University Press.

\bibitem[Tilgner (1999)]{T99}
  \textsc{Tilgner, A.} 1999
  Driven inertial oscillations in spherical shells.
  \textit{Phys. Rev. E} \textbf{59}, 1789--1794.

\bibitem[Zahn (1970)]{Z70}
  \textsc{Zahn, J.-P.} 1970
  Forced oscillations in close binaries.  The adiabatic approximation.
  \textit{Astron. Astrophys.} \textbf{4}, 452--461.

\bibitem[Zhang, Liao \& Earnshaw (2004)]{ZLE04}
  \textsc{Zhang, K., Liao, X. \& Earnshaw, P.} 2004
  On inertial waves and oscillations in a rapidly rotating spheroid.
  \textit{J. Fluid Mech.} \textbf{504}, 1--40.

\end{thebibliography}
\end{document}